\documentclass{ecai} 

\usepackage{caption}
\usepackage{latexsym}
\usepackage{amssymb}
\usepackage{amsmath}
\usepackage{amsthm}
\usepackage{booktabs}
\usepackage{enumitem}
\usepackage{graphicx}
\usepackage{color}
\usepackage{array} 
\usepackage{graphicx} 
\usepackage{cellspace} 
\newcommand{\BibTeX}{B\kern-.05em{\sc i\kern-.025em b}\kern-.08em\TeX}
\usepackage[justification=raggedright,singlelinecheck=false]{caption}

\begin{document}


\begin{frontmatter}



\paperid{123} 




\title{DeepPlantCRE: A Transformer-CNN Hybrid Framework for Plant Gene Expression Modeling and Cross-Species Generalization}

\author[a]{\fnms{Yingjun}~\snm{Wu}\footnote[\dagger]{Contributed equally to this work.}}
\author[a]{\fnms{Jingyun}~\snm{Huang}\footnotemark[\dagger]}
\author[a]{\fnms{Liang}~\snm{Ming}\footnotemark[\dagger]}
\author[a]{\fnms{Pengcheng}~\snm{Deng}}
\author[a]{\fnms{Maojun}~\snm{Wang}\footnote[\ast]
 {Corresponding authors: Maojun Wang (Email:mjwang@mail.hzau.edu.cn) and Zeyu Zhang (Email:zhangzeyu@mail.hzau.edu.cn).}}
\author[a]{Zeyu Zhang\footnotemark[\ast]}

\address[a]{National Key Laboratory of Crop Genetic Improvement, Hubei Hongshan Laboratory, Huazhong Agricultural University, 430070, Hubei, China}




\begin{abstract}
The investigation of plant transcriptional regulation constitutes a fundamental basis for crop breeding, where cis-regulatory elements (CREs), as the key factor determining gene expression, have become the focus of crop genetic improvement research. Deep learning techniques, leveraging their exceptional capacity for high-dimensional feature extraction and nonlinear regulatory relationship modeling, have been extensively employed in this field. However, current methodologies present notable limitations: single CNN-based architectures struggle to capture long-range regulatory interactions, while existing CNN-Transformer hybrid models demonstrate proneness to overfitting and inadequate generalization in cross-species prediction contexts. To address these challenges, this study proposes DeepPlantCRE, a deep-learning framework for plant gene expression prediction and CRE Extraction. The model employs a Transformer-CNN hybrid architecture that achieves enhanced Accuracy, AUC-ROC, and F1-score metrics over existing baselines (DeepCRE and PhytoExpr), with improved generalization performance and overfitting inhibiting. Cross-species validation experiments conducted on gene expression datasets from \textit{Gossypium}, \textit{Arabidopsis thaliana}, \textit{Solanum lycopersicum}, \textit{Sorghum bicolor}, and \textit{Arabidopsis thaliana} reveal that the model achieves peak prediction accuracy of 92.3\%, particularly excelling in complex genomic data analysis. Furthermore, interpretability investigations using DeepLIFT and Transcription Factor Motif Discovery from the importance scores algorithm (TF-MoDISco) demonstrate that the derived motifs from our model exhibit high concordance with known transcription factor binding sites (TFBSs) such as MYR2, TSO1 in JASPAR plant database, substantiating the potential of biological interpretability and practical agricultural application of DeepPlantCRE.
\end{abstract}

\end{frontmatter}

\textbf{Code} : https://anonymous.4open.science/r/DeepPlantCRE-23D5/


\begin{figure}[h]
    \centering
    \includegraphics[width=0.5\textwidth]{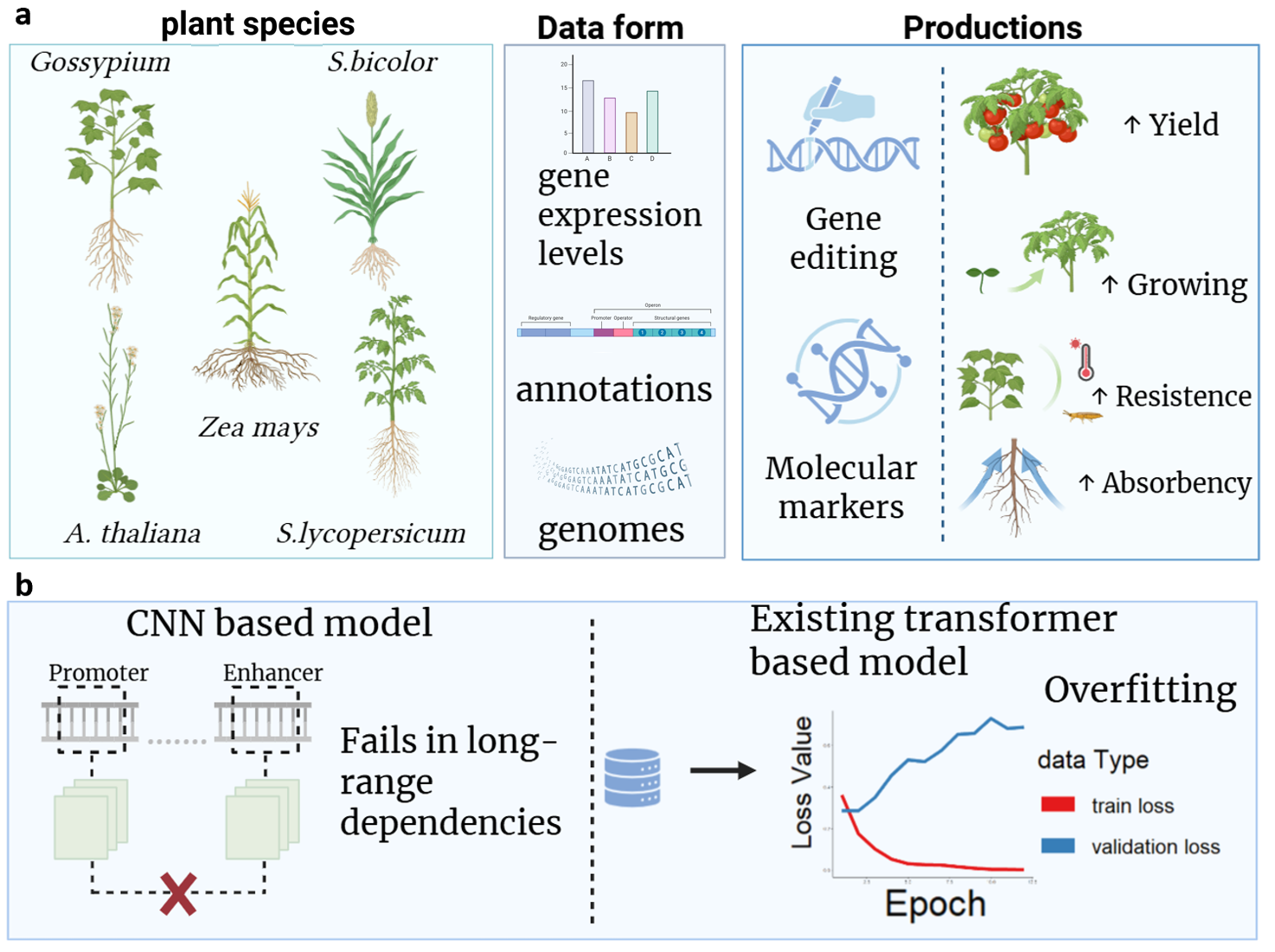}
    \caption{\textbf{a.} Used plant species and data form and future application directions. Species selection covered key plants such as \textit{Gossypium}, \textit{Arabidopsis thaliana}, \textit{Solanum lycopersicum}, \textit{Sorghum bicolor}, \textit{Arabidopsis thaliana}, \textit{Zea mays}. The data module integrates gene expression levels, gene annotations, and genomes. Future applications will focus on gene editing technology optimization, yield improvement, resistance breeding and efficient utilization of resources. \textbf{b.}  Limitations of existing models. The long-range dependence modeling of CNN is insufficient, and Transformer is easy to overfit. The line chart in the figure is based on the training results of the PhytoExpr \cite{li2024modeling} model on the experimental data, showing that the overfitting phenomenon occurred with a significant difference between the training loss and the validation loss.
}
\label{fig:enter-label}
\vspace{12pt}
\end{figure}

\section{Introduction}
The genetic improvement of plant traits depends on the accurate identification and functional elucidation of cis-regulatory elements (CREs). CREs regulate the spatiotemporal expression of genes by binding specifically to transcription factors (TFs), which in turn affects plant growth, stress resistance and other key traits 
 \cite{adamo2021plant,rodriguez2017engineering}. By elucidating the function of CREs, efficient molecular markers can be designed  
 \cite{muller2019impact} and targeted optimization of gene editing technologies can be directed, thereby accelerating the process of crop improvement. However, due to the complexity of plant genome structure, especially the ambiguity of functional characteristics of non-coding regions, it is difficult for traditional methods to systematically identify and resolve the distribution and mechanism of regulatory elements.

In recent years, deep learning techniques have achieved significant breakthroughs in the field of genomics, demonstrating considerable promise, especially in the recognition of CREs. Convolutional Neural Networks (CNNs), due to their strengths in extracting local sequence features, have been widely employed for CRE identification from DNA sequences \cite{alipanahi2015predicting}.
The Transformer architecture is introduced due to its excellent global context modeling ability, which significantly improves the prediction accuracy of the model for long-distance control elements \cite{ashish2017attention,peleke2024deep}.

Although the performance of these methods on single-species datasets has been validated 
 \cite{li2022plants}, the cross-species generalization ability and versatility of existing models are still limited 
 \cite{jung2021autonomous}. Although the model based on the single CNN architecture performs well in local feature extraction 
 \cite{pinto2021fcslib}, it has significant limitations in dealing with long-distance dependencies, especially in the prediction of the interaction between the distal enhancer and the promoter, which is far less effective than the model of global modeling. In addition, existing models based on CNN or Transformer architectures, such as PhytoExpr \cite{li2024modeling}, have good performance on specific datasets, but are prone to overfitting in the case of limited data volume or complex features, resulting in limited generalization ability of models 
 \cite{govers2024apparent,nguyen2023modelling}.

To overcome these limitations, we propose DeepPlantCRE, a deep learning framework based on a hybrid architecture of Transformer and CNN, aiming to address the challenges of cross-species applicability in existing methods. To balance model capacity and generalization ability, we introduce regularization strategies: embedding batch normalization after convolutional layers, and implementing learning rate scheduling during training, effectively suppressing overfitting. We evaluate the model's performance in five representative plant species: cotton (\textit{Gossypium}), maize (\textit{Zea mays}), tomato (\textit{Solanum lycopersicum}), \textit{Arabidopsis thaliana}, and sorghum (\textit{Sorghum bicolor}) through large-scale cross-species cross-validation experiments. The model achieves an accuracy of up to 92.30\%, demonstrating its extremely strong cross-species generalization ability.

In order to systematically evaluate the performance advantages of the DeepPlantCRE model and the effectiveness of its architecture design, comprehensive ablation experiments and parameter sensitivity analysis were also carried out. Experimental results show that the DeepPlantCRE model performs significantly better than the existing benchmark model in cross-species testing. Specifically, compared with the pure CNN model, the complete Transformer-CNNs hybrid architecture improves the prediction accuracy by 1.6\% -3.8\% and the equilibrium F1-score by 1.9\%-3.0\%, which not only confirms the accuracy advantage of the model in the gene expression prediction task but also the enhancement of cis-regulatory element (CRE) identification task. The ablation study further revealed the differential contribution of each module to the performance of the model, among which the multi-head self-attention mechanism significantly improved the modeling ability of long-range dependence, the CNN module with local feature extraction effectively captured the conservative sequence patterns, and the hierarchical feature fusion mechanism optimized the integration efficiency of multi-scale features.

In the hyperparameter sensitivity experiment, we found that the model performance showed good robustness to key parameters such as CNN layers, convolutional kernel size, and learning rate. Experimental results show that when the size of the convolutional kernel is controlled in the range of 5 - 16 bases, the initial learning rate is set in the range of 1e-5 to 1e-3, and the number of convolutional layers is in the range of 4 - 8, the model can not only maintain high prediction accuracy 
 (fluctuation amplitude < 0.06), but also ensure the stability of the training process 
. These findings provide an important parameter optimization basis for the application of deep learning methods in the field of genome cis regulatory element identification, and also lay a theoretical foundation for the engineering deployment of subsequent models.

The contributions of this article are as follows:

1 ) A Transformer-CNN hybrid modeling framework is proposed, which integrates the multi-head self-attention mechanism and convolutional neural network to solve the problem of insufficient long-range regulation modeling and cross-species overfitting.

2 ) Verify the cross-species performance advantage: Among the 5 different plant species, the model accuracy reaches the highest 92.3\%, and the AUC-ROC and F1-score achieves the highest 97.6\% and 92.0\%, respectively.

3 ) Reveal biological interpretability: Interpretability analysis is used to identify conserved regulatory motifs (such as ERF/DREB), and their sequence characteristics are highly consistent with the known TFBS in the database, confirming their biological reasoning potential and production application value.

The structure of this paper is organized as follows: the Related Work section reviews the current advanced applications of machine learning and deep learning techniques in transcriptional regulation; the Proposed Method section introduces the architectural design of the proposed deep learning model, as well as the data processing and training strategies; the Experiments section outlines performance evaluations, the cross-species validation, ablation study, and hyperparameter sensitivity analysis. Finally, the Conclusion section summarizes the primary contributions of this research and outlines potential directions for future studies.


\section{Related Work}

As the development of machine learning and deep learning technologies, their data mining and pattern recognition capabilities have revolutionized the field of gene expression prediction and transcriptional regulation analysis. They can efficiently extract key features from massive gene datasets, construct accurate predictive models, and uncover complex transcriptional regulation relationships, offering productive study with significant opportunities and challenges. Consequently, many research teams have applied these methods actively, achieving substantial progress in this field.

\subsection{Study based on machine learning}

Traditional machine learning methods play an important role in gene expression prediction and transcriptional regulation research. The use of information gain feature selection and logistic regression classification methods to construct models has improved the sensitivity and accuracy of RNA sequencing in identifying differentially expressed genes (DEGs) \cite{wang2018rna}. Rychel et al. used unsupervised machine learning methods (Independent Component Analysis, ICA) to modularize transcriptome data of Bacillus subtilis and quantitatively describe the regulatory activities of gene expression under various conditions \cite{rychel2020machine}. Based on the random forest model, Zhou et al. analyzed the cis-regulatory elements in the promoter region of maize genes and predict their expression for response to heat and cold stress \cite{zhou2022prediction}. Wu et al. proposed to combine tsCUT\&Tag and ATAC-seq data with machine learning techniques using LSTM, TCN, and SVM to assist in predicting transcription factor binding sites in different plant tissues \cite{wu2022cost}. The use of CatBoost algorithm and SHAP interpretation technique have achieved quantitative prediction of TFCR (transcription factor binding site cluster region) complexity and gene expression synchrony \cite{ouyang2024developmental}. Although these methods are efficient in small and medium-sized data, they are limited by feature engineering dependencies and the ability to analyze regulatory complexity.

\subsection{Study of deep learning models}

Deep learning technology has driven breakthroughs in this field through automatic feature extraction and complex pattern modeling. The self-attention mechanism of Transformer architecture is used to mine remote regulatory interactions in sequences \cite{wang2024deepcba}. A model based on interpretable CNNs decodes the cis-regulatory elements of the whole tomato genome through feature visualization technology \cite{akagi2022genome}. Based on the attention mechanism, the parallel prediction of multiple transcription factor binding sites was realized, and the influence of DNA shape characteristics was analyzed \cite{yan2022plantbind}. For large-scale cis-modulation studies, a CNN Transformer multi-task framework was proposed, which significantly improved the accuracy of cis-regulated mutation prediction of 600,000 genes through stacked architecture and hyperparameter optimization \cite{li2024modeling}. Combined with Convolutional Neural Network (CNN) and Bidirectional Long Short-Term Memory Network (BiLSTM), the prediction accuracy of maize gene expression model was significantly improved \cite{avsec2021effective}. However, the existing deep learning models are not interpretable enough to analyze the regulatory mechanism, and the ability to generalize across species is limited by the difference in sequence conservatism.

In this paper, we propose the DeepPlantCRE framework, which adopts the Transformer-CNN hybrid architecture to capture the regulatory mode from local to global for precise prediction on gene expression level.

\section{Proposed Method}
This study proposes a hybrid processing architecture based on lightweight 1D convolutional networks and transformers. After dimensional alignment, the input sequence passes through transformer layers to capture long-range dependencies, then undergoes multiscale feature extraction via hierarchical residual convolution modules to reduce the number of parameters and finally its features are refined for classification. The detailed workflow is as follows.

\captionsetup{skip=10pt}
\begin{figure*}
    \centering
    \includegraphics[width=0.8\linewidth]{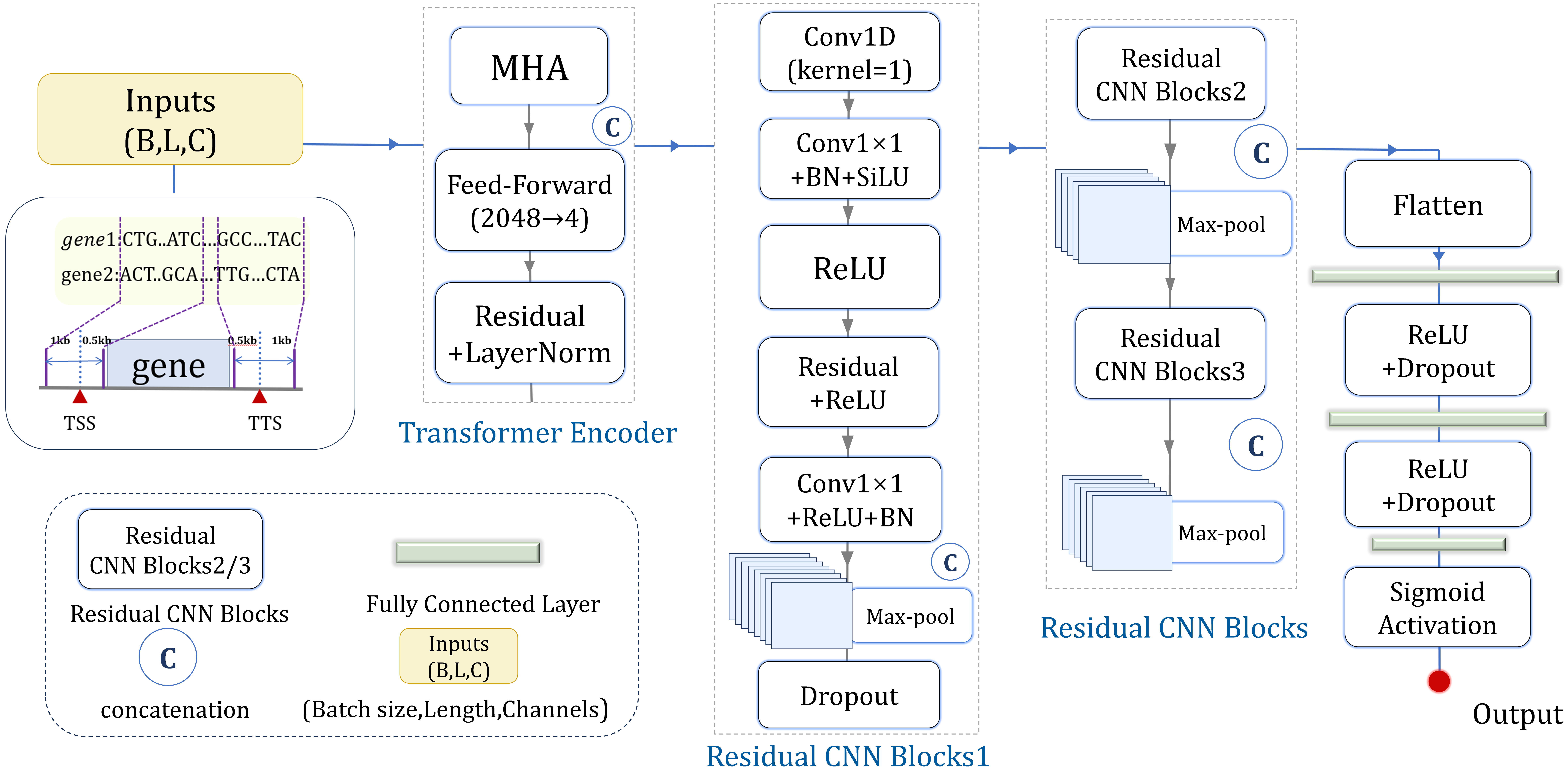}
    \caption{Model structure overview: The model consists of an optional Transformer encoder and three cascaded convolution residual blocks, and is finally connected to a binary classification decoder. All modules share the same input embedding (unique hot encoding). The Transformer encoder is responsible for extracting globally dependent features. The three-segment residual convolutional blocks gradually increase channels, compress the base sequence and mine the local basis order. Finally, the features become stable and then enter to the fully connected classification head to generate the binary classification probability of the current input sequence.
    }
    \label{fig:enter-label}
    \vspace{8pt}
\end{figure*}
\subsection{Input Representation and Preprocessing}
\label{sec:input}
When presented with a DNA sequence $S = (s_1, s_2, ..., s_L)$, where $s_i \in \{A,C,G,T,N\}$, we construct a matrix encoded with one hot.
\begin{equation}
X \in \{0,1\}^{L \times D}, \quad X_{i,j} = 
\begin{cases} 
1, & \text{if } s_i \text{ is the } j\text{-th base type} \\
0, & \text{otherwise}
\end{cases}
\end{equation}
where $L$ is the sequence length of DNA and $D$ (the one-hot vector dimension) equals the number of standard base types.
\subsection{Transformer Encoder Layer}
\label{sec:transformer}
We find that it is beneficial to use different learned linear projections to map queries, keys, and values to dimensions $d_q$, $d_k$, and $d_v$, respectively, rather than applying a single attention function $h$ times to the same $d$-dimensional queries, keys, and values. Then, on each projected version of queries, keys, and values, we execute the attention function in parallel, generating $d$-dimensional output values. These outputs are concatenated and projected again to obtain the final values. Multi-head attention allows the model to jointly attend to information from different representation subspaces at different positions; for a single attention head, averaging would suppress this. To enable the model to respond in parallel to different subspaces and different positions, the paper proposes projecting queries, keys, and values $h$ times, respectively, resulting in dimensions of $d_q$, $d_k$, and $d_v$. Then multiple groups of $(Q_i, K_i, V_i)$ are calculated in parallel, and their attention outputs are concatenated and projected. The encoder stack contains $N=1$ layer with one multi-head attention block followed by a position-wise feed-forward network \cite{vaswani2017attention}.

 (i) Multi-Head Attention (1 head). 
  \begin{equation}
  \text{head}_i = \text{Softmax}\left(\frac{Q_iK_i^T}{\sqrt{d_k}}\right)V_i, \quad d_k = d_v = 4
  \end{equation}
  
 (ii) Feed-Forward Network. 
Following each attention sub-layer, we apply a two-layer, position-wise feed-forward network (FFN) with ReLU activation. 
\begin{equation}
  \text{FFN}(U) = W_2(\text{ReLU}(W_1U + b_1)) + b_2
  \end{equation}
\begin{equation}
W_1 \in \mathbb{R}^{4 \times 2048}, W_2 \in \mathbb{R}^{2048 \times 4}
\end{equation}

Model generates output after Transformer processing. 
\begin{equation}
H^{(i+1)} = \text{LayerNorm}(H^{(i)} + \text{FFN}(H^{(i)}))
\end{equation}
\subsection{Convolutional Residual Blocks}
\label{sec:conv}
Although self-attention captures the global context, local motifs in biological sequences are crucial. Therefore, we stack three one-dimensional convolutional residual blocks, each is followed by pooling and dropout, to extract hierarchical local patterns from the transposed transformer output, $U \in \mathbb{R}^{D \times L}$. 

(i) CNN Block Operations. 
\begin{align}
A_1 &= \text{ReLU}(\text{Conv1D}(U)) \quad (\text{kernel}=8) \label{eq:conv1} \\
A_2 &= \text{ReLU}(\text{Conv1D}(A_1)) \quad (\text{kernel}=8) \label{eq:conv2} \\
R_1 &= \text{Conv1D}(U) \quad (\text{kernel}=1) \label{eq:res1} \\
U^{'} &= \text{ReLU}(\text{BN}(A_2) + R_1) \label{eq:resadd1} \\
U_f &= \text{MaxPool}_p(\text{Dropout}(U^{'})) \label{eq:pool1}
\end{align}
where $U$ represents the input feature map of the current residual block, with $C$ input channels corresponding to the Transformer’s embedding dimension and length $L$ matching the original DNA sequence. Given the primary convolution output $A_1$ and the secondary output $A_2$, the fused feature $U^{'}$ is produced, and after pooling the final output $U_f$ is obtained. The subsequent blocks in the network follow a similar pattern, with channel dimensions evolving from 64, to 128, and then to 32.

 (ii) Classification Head. 
\label{sec:head}
We then apply three dense layers with ReLU activations and dropout to produce the final output via a sigmoid activation for binary classification or probability regression. The final prediction is computed as
\begin{equation}
\hat{y} = \sigma\left(W_3 \cdot \text{ReLU}\left(W_2 \cdot \text{ReLU}\left(W_1 f\right)\right)\right)
\end{equation}
where $f = \text{Flatten}(U) $ and $W_1 \in \mathbb{R}^{D_1 \times (C \times L')}$, $W_2 \in \mathbb{R}^{D_2 \times D_1}$, and $W_3 \in \mathbb{R}^{D_3 \times D_2}$, we employ a sigmoid activation function defined by $\sigma(z) = 1/(1+e^{-z})$, which is utilized to introduce non-linearity into the model. Here, the flatten operation transforms the tensor $U$ into a 2D matrix $f$, which serves as the input to the subsequent fully connected layers. These layers, parameterized by the weight matrices $W_1$, $W_2$ and $W_3$, facilitate the learning of hierarchical representations through the network.

(iii) Loss function. Dropout is applied after each hidden layer to mitigate overfitting. After obtaining the single-valued probability output of model through the sigmoid function, we use BCEWithLogitsLoss for numerical stability.
\begin{equation}
\mathcal{L} = -\frac{1}{N} \sum_{i=1}^N \left[ y_i \log \sigma(\hat{y}_i) + (1 - y_i) \log(1 - \sigma(\hat{y}_i)) \right]
\end{equation}
In the model evaluation process, $y_{i}$ denotes the true labels, which are binary values of either 0 or 1, while $\hat{y}_i$ represents the model predicted probabilities produced by the sigmoid output layer. Meanwhile, $N$ refers to the sample size. 
\subsection{Optimization Strategy}
 (i) Adam optimizer with learning rate $\eta=10^{-4}$. 
  \begin{equation}
  \theta_{t+1} = \theta_t - \eta \frac{\hat{m}_t}{\sqrt{\hat{v}_t} + \epsilon}
  \end{equation}
  
 (ii) Scheduling of the learning rate. We use the ReduceLROnPlateau. When the monitoring metric does not decline for consecutive epochs, multiply the current learning rate by a decay factor. 
 
 (iii) Early Stop. Model training will be terminated if validation loss seldom changes for n epochs (n = 10). 
 
 (iv) Regularization. In terms of regularization, we adopt dropout with a probability of $p=0.25$ after each dense and convolutional layer to randomly deactivate neurons, thus enhancing the generalization ability of model. Moreover, batch normalization is applied in residual branches to normalize inputs and accelerate convergence. 

\subsection{Computational Complexity}
\label{sec:complexity}
The time complexity combines Transformer and CNN components. 
\begin{equation}
\mathcal{O}\left(N\left(L^2d + 3Ld^2 + 2Ldd_{\text{ff}}\right)\right) + \sum_{j=1}^3 \mathcal{O}\left(\left(C_{j-1}C_j + C_j^2\right)kL_j\right)
\end{equation}
where $N=1$ for the number of Transformer layers, which is designed to capture complex patterns in the data. For the embedding dimensions, we choose $d=4$ and $d_{\text{ff}}=2048$, with the aim of balancing computational efficiency and representation power. CNN channels are specified as $C_j=\{64,128,32\}$, which allows the model to extract features at different levels of abstraction and complexity. We fix the convolution kernel size at $k=8$ to effectively capture local context while maintaining a reasonable receptive field. The length of the sequence at each layer is denoted by $L_j$, which varies depending on the specific layer and its role in the network. 

\section{Experiments}

In this section, we carried out a series of experiments to explore 4 crucial questions list below:

\begin{itemize}
    \item \textbf{Q1:} How does our model DeepPlantCRE perform on 5 different species datasets?
    \item \textbf{Q2:} What is the performance of DeepPlantCRE when conducted in cross-species experiments?
    \item \textbf{Q3:} How do hyperparameters impact the performance of model?
    \item \textbf{Q4:} How does each module of our model contribute to expression level prediction?
\end{itemize}

\subsection{Experiment Settings}

For this part, we applied our model DeepPlantCRE to diverse crop species datasets to exhibit the model effectiveness.

\textbf{Datasets.} The detailed introduction to datasest structure in this project are as follows.

\textbf{8 diploid \textit{Gossypium} species:} The genome files of \textit{G. arboreum} (A2), \textit{G. anomalum} (B1), \textit{G. sturtianum} (C1), \textit{G. raimondii} (D5), \textit{G. stocksii} (E1), \textit{G. longicalyx} (F1), \textit{G. bickii} (G1), \textit{G. rotundifoliumm} (K2) including DNA sequences of all 13 chromosomes, annotation files of 41,000, 40,488, 41,308, 40,281, 40,295, 41,633, 40,857 and 41,590 genes respectively which include the start site and the end site of each gene, and gene expression level files for leaf tissue. 

\textbf{\textit{Arabidopsis thaliana}:} The reference genome file containing DNA sequences for 5 chromsomes, annotation file for 27,655 genes covering the start site and the end site of each gene, and gene expression level flie for leaf tissue. 

\textbf{\textit{Solanum lycopersicum}:} The reference genome file which includes DNA sequences of all 10 chromsomes, annotation file for 34,658 genes, and gene expression level of leaf tissue.

\textbf{\textit{Sorghum bicolor}:} The reference genome that includes DNA sequences about all 12 chromsomes, annotation file of 34,118 genes involving the start site and the end site of each gene, and their corresponding gene expression level in leaf tissue.

\textbf{\textit{Zea mays}:} The reference genome file which includes DNA sequences of all 10 chromsomes, annotation file of 39,757 genes, and gene expression level file for leaf tissue.

All raw sequencing data and genome assemblies of 8 diploid \textit{Gossypium} species with gene annotation files were downloaded from the National Center for Biotechnology Information (NCBI) database \cite{wang2022genomic}.
The reference genomes and gene annotation files of \textit{A. thaliana, S.lycopersicum, S. bicolor and Z. mays} were downloaded from Ensembl Plants database (v52) (plants.ensembl.org).

\textbf{Data Processing.} Specifically, after accessing genomes and gene annotation files, the regions are extracted for each gene ranging from 1kb upstream to 0.5kb downstream of the transcription start site (TSS) and from the 0.5kb upstream and 1kb downstream of transcription termination site (TTS) as a part of the input for model. 

Regarding the gene expression levels for leaf tissue, we got the transcriptome datas of 8 diploid \textit{Gossypium} from NCBI and then dealt RNA sequences with removing adaptors and low-quality base clipping (Trimmomatic) \cite{bolger2014trimmomatic}, aligning clean reads to the reference genome (HISAT2) \cite{kim2015hisat} followed by removing PCR duplicates (SAMtools, -q 20) \cite{li2009sequence} to compute the gene expression levels with StringTie \cite{pertea2015stringtie,you2023regulatory}, then calculated the log-transformed transcript per million values (logMaxTPM). Also, we utilized datas for other 4 plants in previous study that they downloaded short-read transcriptomic data for leaf issue from the National Center for Biotechnology Information (NCBI) Sequence Read Archive (SRA) database using the fasterq-dump and trimmed them by sickle and then aligned the reads to the reference cDNA by Kallisto, thereby acquiring the standardized count of genes per million transcripts (TPM) after results processing with
the tximport package in R \cite{peleke2024deep}. For each species, gene expression levels were categorized into low, medium and high groups based on the distribution of the logMaxTPM values < 25\% marked as -1 (low expression), 25\%-75\% as 0 (medium expression), and > 75\% as 1 (high expression). 

\textbf{Baselines.} Baseline methods include two deep learning models - DeepCRE \cite{peleke2024deep} and PhytoExpr \cite{li2024modeling}.

\textbf{DeepCRE:} It employed Convolutional Neural Network (CNN) technology to analyze plant genomic data, predict gene expression levels, and identify DNA sequences with regulatory functions \cite{peleke2024deep}.

\textbf{PhytoExpr:} It used Convolutional Neural Network (CNN) technology and Transformer layer to predict messenger RNA (mRNA) abundance by analyzing plant cis-regulatory elements \cite{li2024modeling}.

\textbf{Metrics.} We assessed the  predictive performance of deep learning model using accuracy, AUC, and F1-score. 
Accuracy, the ratio of correctly predicted samples to the total, measures overall correctness. The AUC-ROC curve with horizontal axis - False positive rate and vertical axis - True Rate, which can show the capacity of model to distinguish between positive class and negative class. The F1-score is a harmonic mean of precision and recall balancing the two measures.

\textbf{Configurations.} All experiments were deployed and executed on a 64-bit machine equipped with two NVIDIA GPUs (NVIDIA L20, 1440 MHz, 48 GB memory). 

For our model DeepPlantCRE, we employed Adam optimizer with a learning rate of $10^{-4}$. 
Model training was restricted to a maximum of 100 epochs with forward propagation, back-propagation, and parameter updates on training data for each epoch. To avoid overfitting, early stopping is used to halt training if the validation loss does not decrease within 10 consecutive epochs. Additionally, the learning rate was dynamically adjusted, reducing to 10\% of its current value if the validation loss does not minish for 5 consecutive epochs. This dynamic learning rate adjustment strategy aids finer convergence to optimal solutions in the late training phase. Utilizing a \textit{k} fold cross-validation strategy, the best-performing model was generated after all cross-validation folds.

To be specific, as for the Encoding Layer, DeepPlantCRE uses 1 Transformer Layer with an embedding dimension of 4 matching the one-hot encoding of the four DNA nucleotides (A, T, C, G). The Transformer encoder layer also includes a feed-forward neural network with a hidden layer dimension of 2,048, enabling to learn more complex feature representations. A dropout rate of 0.25 is applied to prevent overfitting and boost the generalization ability of model. Besides, the model comprises 6 convolutional layers with the kernel size of 8 for each layer. 3 residual connections are incorporated using $1 \times $1 convolutions to adjust channel numbers to alleviate gradient disappearance in deep networks. About the output, a fully connected layer consisting of 3 linear layers with decreasing neuron numbers eventually output a continuous value representing the predicted gene expression level. The fully connected layer also uses ReLU activation and dropout regularization, with a sigmoid function in the final layer to map outputs to the [0, 1] interval for adapting gene expression classification task.

\subsection{Model Performance (RQ1)}
In order to answer RQ1 to demonstrate the superiority of our model, we trained and tested it on multiple crops datasets. For each species, we employed \textit{k} fold cross-validation for training, with \textit{k} setting as the number of chromosomes of each species under training. In order to make comparison with other reported deep learning models, we also applied these same datasets to other models. Finally, we evaluated the performance of model by analyzing the 3 indicators including Accuracy, AUC-ROC, and F1-score.

From Table \ref {tab:performance of all species}, we get the conclusion that DeepPlantCRE outperforms DeepCRE \cite{li2024modeling} in terms of every indicator. Our model showed a 1.6\%-3.8\% better performance than other reported models in the Accuracy metric and achieved higher AUC-ROC and F1-score about 1.9\%-3.0\% and 1.2\%-2.8\% respectively. This is attributed to the processing order in our model by prioritizing Transformer layer to capture global interaction information, and then employing CNN module to extract local features. Moreover, by integrating CNN with Transformer, DeepPlantCRE enhances the effective features information and better balances long-distance interactions and local relationships of each nucleotide, giving it an edge over PhytoExpr \cite{peleke2024deep}.

\setlength{\intextsep}{10pt} 
\setlength{\abovecaptionskip}{5pt} 
\setlength{\belowcaptionskip}{3pt} 

\begin{table}[h]\large
\centering
\resizebox{0.49\textwidth}{!}{
  \begin{tabular}{cc|c|c|c}
  \hline
  \multicolumn{2}{c|}{Dataset} & \textbf{DeepCRE}     & \textbf{PhytoExpr}     & \textbf{DeepPlantCRE}  \\
  \hline
                        & Accuracy    & 89.3±3.6          & 87.1±2.6 & \textbf{90.9±2.0}   \\
\textbf{\textit{Gossypium arboreum} (A2)}           & AUC-ROC     & 94.7±2.0            & 93.6±2.2 & \textbf{96.6±1.4} \\
                        & F1-score    & 92.2±2.8          & 90.6±2.3 & \textbf{93.4±1.7} \\ \hline
                        & Accuracy    & 85.4±2.0   & 78.3±1.5 & \textbf{86.1±3.0}   \\
\textbf{\textit{Arabidopsis thaliana}}    & AUC-ROC     & 92.5±1.8   & 86.3±0.9 & \textbf{93.0±1.7}   \\
                        & F1-score    & 85.0±1.9 & 76.6±2.4 & \textbf{85.8±2.8} \\ \hline
                        & Accuracy    & 82.6±2.7          & 79.2±3.0   & \textbf{84.3±2.6} \\
\textbf{\textit{Solanum lycopersicum}}    & AUC-ROC     & 87.8±2.4          & 85.2±3.1 & \textbf{90.0±2.3}   \\
                        & F1-score    & 74.7±4.5          & 71.3±5.5 & \textbf{77.9±3.9} \\ \hline
                        & Accuracy    & 79.0±1.9            & 75.7±2.4 & \textbf{80.7±1.6} \\
\textbf{\textit{Sorghum bicolor}}         & AUC-ROC     & 84.4±5.3          & 81.9±2.3 & \textbf{86.2±4.8} \\
                        & F1-score    & 73.5±6.9          & 71.5±5.2 & \textbf{76.1±6.4} \\ \hline
                        & Accuracy    & 80.1±3.9          & 77.9±4.1 & \textbf{83.0±4.5}   \\
\textbf{\textit{Zea mays}}                & AUC-ROC     & 86.7±5.1          & 85.2±4.0   & \textbf{90.0±4.1}   \\
                        & F1-score    & 78.0±5.8            & 76.2±5.1 & \textbf{81.2±6.1} \\ \hline
  \end{tabular}

}
\caption{The performance of different deep learning models on datasets of 5 diverse crops. Model prediction results (average ± standard deviation) with Accuracy (\%) and AUC-ROC (\%) and F1-score (\%). The best scores are displayed in bold.} 
\label{tab:performance of all species}
\end{table}

To further explore the performance of DeepPlantCRE in handling species with complex and large-scale genomes and its ability of distinguishing variety characteristics for gene prediction, we apply DeepPlantCRE for 8 diploid \textit{Gossypium} datasets differing significantly from each other in terms of the size of genome (ranging from 750 Mb to 2,444 Mb), structure of chromosomes and the phenotypic traits. For instance, in the long-fiber cotton (A2), certain genes related to fiber development show high expression levels and are governed by specific gene expression networks, while some key genes of short-fiber cotton (K2) are suppressed with the insertion of Transposable Elements \cite{wang2022genomic}. 

The prediction results in Table \ref {tab:performance of 8 Gossypium} shows that our model demonstrates superior performance in accuracy, AUC-ROC and F1-score, outperforming other reported models. It effectively identifies the unique character among varieties by employing transformer layer to consider long-range interactions within DNA sequences. What is more, there is another strategy between the two modules -  batch normalization process for stabilizing data. Moreover, residual connections settings in our model aid in addressing the gradient vanishing problem during model training. Therefore, DeepPlantCRE has more comprehensive and outstanding advantages for prediction.

\begin{table}[h]
\large
\centering
\resizebox{0.49\textwidth}{!}{
  \begin{tabular}{cc|c|c|c}
  \hline
  \multicolumn{2}{c|}{Dataset} & \textbf{DeepCRE}     & \textbf{PhytoExpr}     & \textbf{DeepPlantCRE}  \\
  \hline
                        & Accuracy    & 89.3±3.6          & 87.1±2.6 & \textbf{90.9±2.0}   \\
\textbf{\textit{G. arboreum} (A2)}           & AUC-ROC     & 94.7±2.0            & 93.6±2.2 & \textbf{96.6±1.4} \\
                        & F1-score    & 92.2±2.8          & 90.6±2.3 & \textbf{93.4±1.7} \\ \hline
                        & Accuracy    & 89.1±2.8          & 86.9±5.0   & \textbf{91.4±2.3} \\
\textbf{\textit{G. anomalum} (B1)}           & AUC-ROC     & 94.7±2.2          & 93.7±2.9 & \textbf{96.7±1.4} \\
                        & F1-score    & 91.3±2.5          & 89.6±4.6 & \textbf{93.3±2.0}   \\ \hline
                        & Accuracy    & 90.4±3.6          & 88.5±3.3 & \textbf{91.5±3.1} \\
\textbf{\textit{G. sturtianum} (C1)}           & AUC-ROC     & 95.7±2.5          & 94.4±2.5 & \textbf{97.1±2.0}   \\
                        & F1-score    & 92.3±3.0            & 90.7±2.8 & \textbf{93.2±2.6} \\ \hline
                        & Accuracy    & 88.2±2.8          & 86.9±2.7 & \textbf{91.9±2.3} \\
\textbf{\textit{G. raimondii} (D5)}           & AUC-ROC     & 94.2±2.5          & 94.0±2.3   & \textbf{97.2±1.4} \\
                        & F1-score    & 89.8±3.2          & 88.9±2.7 & \textbf{93.0±2.6}   \\ \hline
                        & Accuracy    & 89.5±3.8          & 87.9±2.8 & \textbf{92.2±2.6} \\
\textbf{\textit{G. stocksii} (E1)}           & AUC-ROC     & 94.9±2.1          & 94.4±1.8 & \textbf{97.4±1.4} \\
                        & F1-score    & 91.8±3.3          & 90.6±2.5 & \textbf{94.0±2.3}   \\ \hline
                        & Accuracy    & 91.1±2.4          & 88.8±2.7 & \textbf{93.1±2.0}   \\
\textbf{\textit{G. longicalyx} (F1)}           & AUC-ROC     & 96.4±1.4          & 95.9±1.3 & \textbf{98.0±1.0}     \\
                        & F1-score    & 92.9±2.2          & 91.3±2.3 & \textbf{94.6±1.7} \\ \hline
                        & Accuracy    & 89.4±2.8          & 87.3±2.4 & \textbf{91.3±2.8} \\
\textbf{\textit{G. bickii} (G1)}           & AUC-ROC     & 94.7±2.0            & 94.0±1.7   & \textbf{96.4±1.2} \\
                        & F1-score    & 91.8±2.2          & 90.3±1.9 & \textbf{93.3±2.3} \\ \hline
                        & Accuracy    & 86.3±2.2          & 85.2±2.5 & \textbf{89.5±1.8} \\
\textbf{\textit{G. rotundifoliumm} (K2)}           & AUC-ROC     & 92.9±1.4          & 92.6±2.4 & \textbf{96.0±1.1}   \\
                        & F1-score    & 89.6±1.9          & 88.8±2.2 & \textbf{92.0±1.6}   \\ \hline
  \end{tabular}

}
\caption{Comparision of the Accuracy (\%) and AUC-ROC (\%) and F1-score (\%) across different models based on 8 diploid \textit{Gossypium} datasets.}
\label{tab:performance of 8 Gossypium}
\end{table}

\subsection{Transfer Learning Study (RQ2)}

For RQ2, we conduct cross-species experiments by using one-species-trained model to test other species to access the generalization ability of our models.

\begin{figure}
    \centering
    \includegraphics[width=0.7\linewidth]{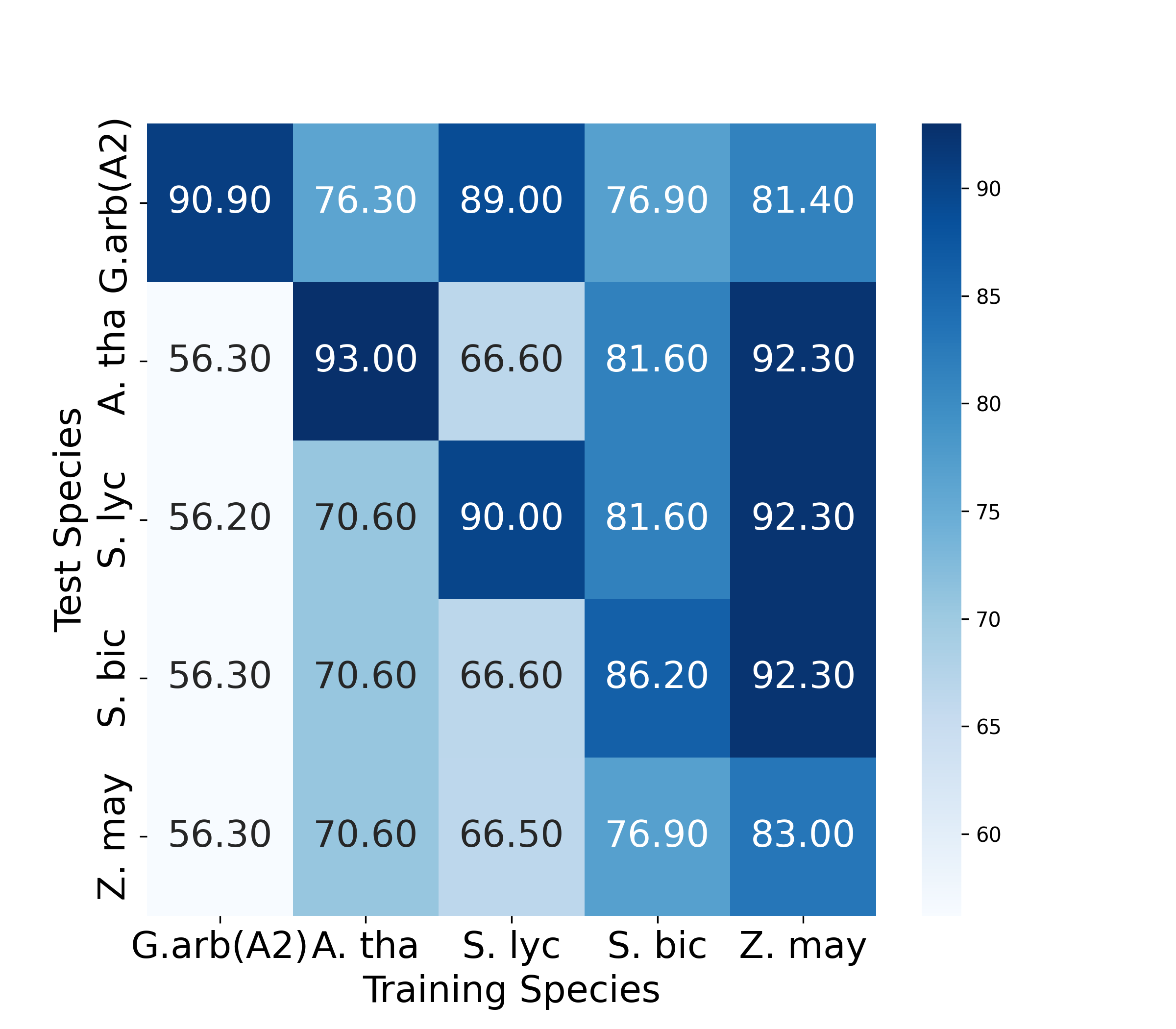}
    \caption{Cross-species experiments on DeepPlantCRE. The x-axis represents the trained species with the y-axis represents the other species being tested. The values show model prediction accuracy.}
    \label{fig:cross-species}
\end{figure}

From the heatmap of accuracy result in figure \ref{fig:cross-species}, we find that the model of \textit{Zea Mays} has the best cross-species performance with the highest prediction accuracy of 92.30\% among other species datasets. Conversely, the model of \textit{Gossypium} performs worst with the accuracy of 56.30\% on cross-species situation. This indicates that model cross-species prediction is closely linked to evolutionary and phylogenetic relationships of species, with greater evolutionary distance generally leading to weaker generalization. On the basis of the first condition, because the genes dataset of \textit{Zea Mays} has 39,757 records more than other species with close relationship, we infer that model tends to have better cross-species effectiveness for species with larger genes datasets for training.

\subsection{Hyperparameter Sensitivity Analysis (RQ3)}

For this section, we altered the hyperparameters, including CNN layers, CNN Kernel Size and Learning Rate to observe the associated model performance changes to give an answer to RQ3. 

\begin{figure*}
    \centering
    \includegraphics[width=0.8\linewidth]{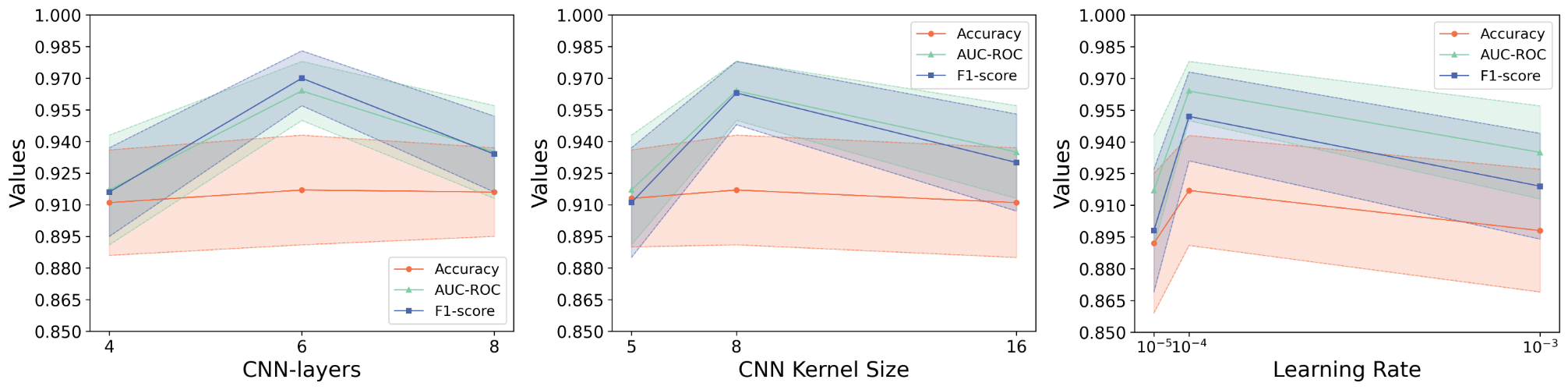}
    \caption{Hyperparameter Sensitivity Study on \textit{G. ano} (B1) dataset, including the CNN Layers, CNN Kernel Size and Learning Rate.}
    \label{fig:hyperparameter}
\end{figure*}

By changing the number of CNN layers from to 4 to 6 to 8, the results on the left side of figure \ref{fig:hyperparameter} shows that model performance displays slight fluctuation but remain roughly constant. About the changes of CNN Kernel Size, model shows a relatively stable trend with the best performance at the kernel size of 8 presented in the center of Figure \ref{fig:hyperparameter}. In terms of learning rate, after testing values of $10^{-3}$, $10^{-4}$, $10^{-5}$, we find model performs fairly consistent with optimal results achieves at $10^{-4}$ as shown in the third picture of figure \ref{fig:hyperparameter}. Therefore, DeepPlantCRE shows low sensibility to the hyperparameter variations.

\subsection{Ablation Study (RQ4)}

To figure out the effectiveness for each individual module in our model, we carry out the Ablation Study on two different species datasets - \textit{Gossypium anomalum} (B1) and \textit{Solanum lycopersicum}. Initially, after removing the Transformer layer, we conduct model training and testing, and then find its accuracy, AUC-ROC and F1-Score declines and becomes more unstable, which probably is caused by overfitting to specific data patterns or highly sensitivity to input-data variations. On the contrary, the model combined Transformer with CNN has better robustness and consistency on both two datasets which have different characteristics of DNA sequences.
\begin{figure}
    \centering
    \includegraphics[width=1\linewidth]{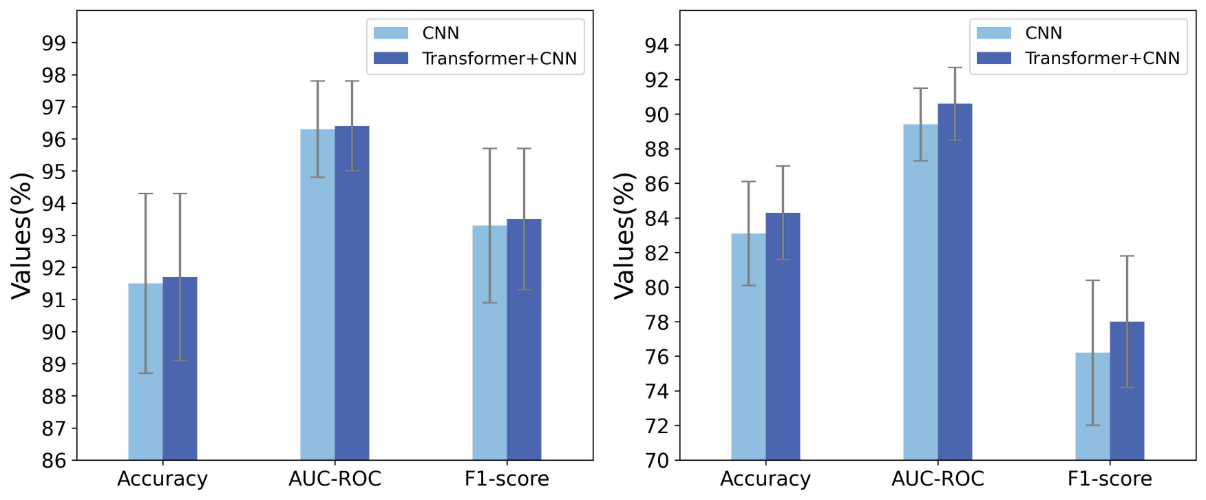}
    \caption{Ablation Study on \textit{G. ano} (B1) and \textit{Solanum lycopersicum} datasets. Light blue: model only based on CNN, dark blue: model integrated Transformer and CNN.}
    \label{fig:ablationstudy}
\end{figure}

\subsection{Case Study}

To verify the scientificity during prediction process of our model and find out its practical value in production, we did this further experiment from a biological perspective.

Firstly, in order to gain insights into how DeepPlantCRE makes prediction on gene expression with the information of nucleotides in DNA sequences, we adopt DeepLIFT \cite{shrikumar2017learning} to interpret the weights of model to get contribution weight for each nucleotide, and further extracts significant motif fragments by TF-MoDISco algorithm \cite{shrikumar2018technical}. These two steps effectively reveals the internal mechanisms of model prediction on gene expression based on DNA sequences.

\begin{figure}
    \centering
    \includegraphics[width=0.9\linewidth]{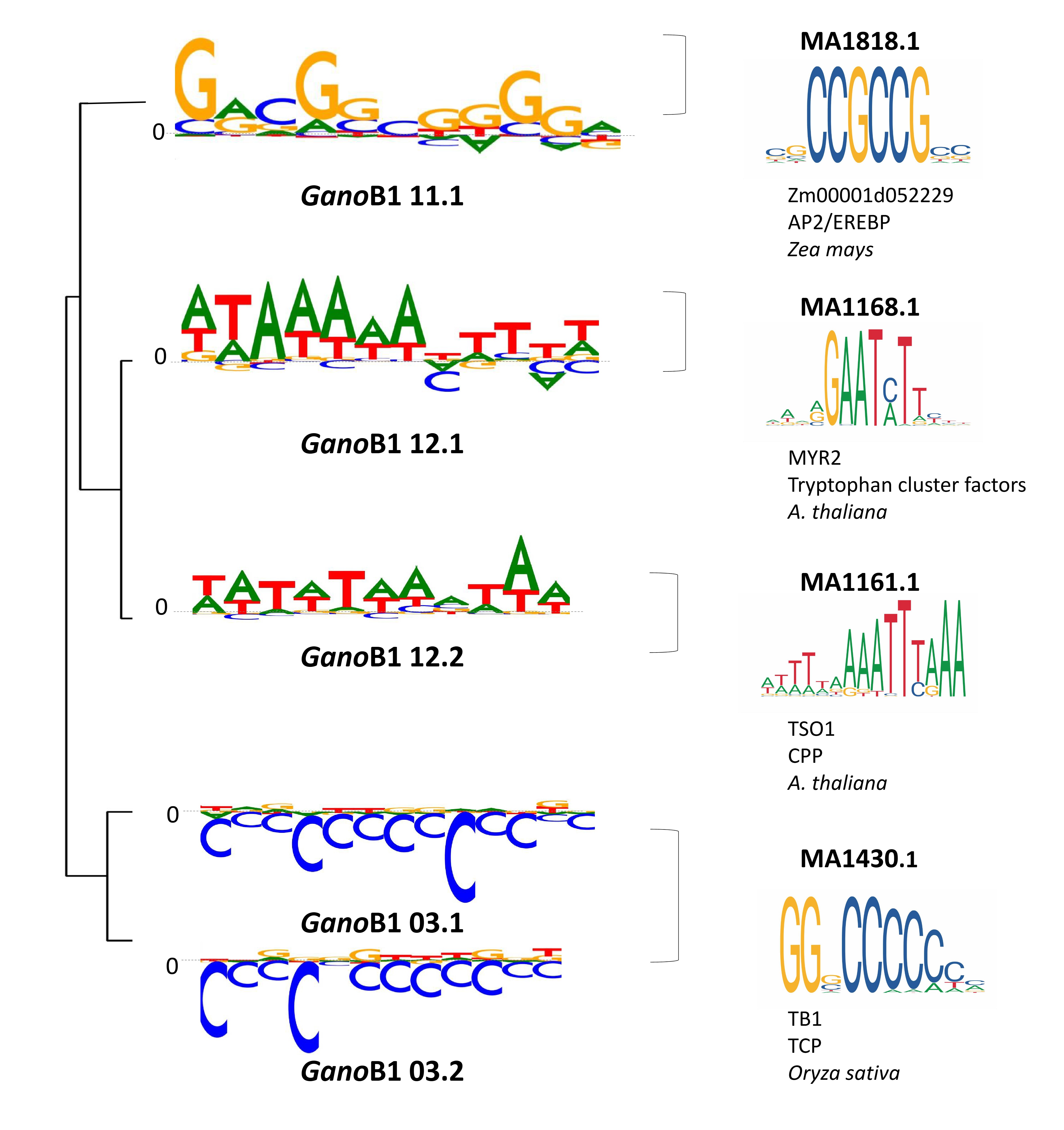}
    \caption{ Several representative groups of all motifs clustering. The logos under x-axis indicate negative contribution scores. }
    \label{fig:clusterPWM}
    \vspace{8pt}
\end{figure}

What is more, comparing these motif segments with the known transcription factor binding sites (TFBSs) in the JASPAR Database \cite{castro2022jaspar} by MEME tools \cite{gupta2007quantifying}, we find most of them matches the known TFBS in a significant way, which indicates that key motif fragments identified by our model can align with actual TFBSs and hold practical biological significance, thereby validating the reliability and scientificity of the outcome from our model. For instance, one motif fragment (at the left of firt row in Figure \ref{fig:clusterPWM}) exhibits high similarity at the sequence level (p value = 9.61e-06 and e value = 1.83e-02) to known TFBS Zm00001d052229 in \textit{Zea Mays} \cite{tu2020reconstructing} which belongs to the AP2/EREBP class. It is one of the largest transcription factor families playing a key role in modulating gene expression during abiotic stresses and also impacting plant growth and development \cite{maghraby2024genome}.

Additionally, using R package motifstack to cluster all the effective motif fragments based on homology information to divide them into different groups \cite{ou2018motifstack}, we picked up several representative groups in figure \ref{fig:clusterPWM} covering negative and positive effects on gene expression to make a comprehensive evaluation. We named each motif fragment according to species and function. The first part represents the species by abbreviation followed by genome information, and next part represents its positive or negative effect on gene expression (1 indicates high expression, 0 indicates low expression), with clustering group ID and the serial number in group following. We found that each group matches the known TFBS in JASPAR such as Zm00001d052229, MYR2, TSO1, and TB1. Specifically, MYR2 and MYR1 negatively regulate GA20ox2 expression, reducing bioactive GA levels, thereby inhibiting flowering and organ elongation \cite{zhao2011arabidopsis}. TSO1 simultaneously regulates cell division and meristem organization in \textit{Arabidopsis} \cite{song2000regulation}. OsTB1 in rice inhibits the growth of axillary buds by regulating the cytokinin and auxin \cite{takeda2003ostb1}.

The explainable results from our model can help pepole do gene editing and breeding of superior varieties. The analysis above reveals that our model is practical and valuable in production.

All the exploration proves that excavated features by our model have high similarity to cis-regulatory elements (CREs) during the gene expression prediction which are strongly associated with the real gene transcriptional regulatory mechanism in plants. So it can contribute to production in the future by predicting CREs of plants just based on DNA sequences.

\section{Conclusion}

In this project, we solved the problem of predicting gene expression levels just based on DNA sequences within the promoter region using our deep learning model DeepPlantCRE integrating Transformer with CNN. Besides, our model reaches higher precision considering the long-distance interaction between nucleotide and the local relationship of them. Also, by interpreting the weights of model and extracting the important motif fragments according to contribution, we obtain a large number of motifs that play an crucial role in gene expression, which is helpful to breeding project of excellent varieties and crops resistance enhancement. In the future, we consider to use multi-omics data to train the deep learning model to achieve more comprehensive gene expression levels prediction and provide better convenience for crop production.


\bibliography{ref}

\end{document}